\documentclass{article}
\usepackage[square,numbers]{natbib}
\usepackage{graphicx} 
\usepackage{pseudo} 
\usepackage{tcolorbox}
\tcbuselibrary{skins,theorems} 
\pseudoset{indent-length=1.25em}
\usepackage{amsmath}
\DeclareMathOperator*{\argmax}{argmax}

\title{Automated Discovery of Improved Constant Weight Binary Codes}
\author{
    Christopher D. Rosin \\
    https://constructive.codes \\
    christopher.rosin@gmail.com
}
\date{}

\begin{document}

\maketitle

\begin{abstract}
A constant weight binary code consists of $n$-bit binary codewords, each with exactly $w$ bits equal to 1, such that any two codewords are at least Hamming distance $d$ apart.  $A(n,d,w)$ is the maximum size of a constant weight binary code with parameters $n,d,w$.  We establish improved lower bounds on $A(n,d,w)$ by constructing new larger codes, for 24 values of $(n,d,w)$ with $6 \leq d \leq 18$ and $18 \leq n \leq 35$.  The improved lower bounds come from two strategies.  The first is a tabu search that operates at the level of bit swaps.  The second is a novel greedy heuristic that repeatedly chooses the candidate codeword that maximizes a randomly-scored histogram of distances to previously-added codewords.  These strategies were proposed by CPro1, an automated protocol that generates, implements, and tests diverse strategies for combinatorial constructions.
\end{abstract}

\section{Introduction}
A \textit{constant weight binary code} of length $n$, weight $w$, and distance $d$ consists of codewords in $\{0,1\}^n$, each having Hamming weight $w$ (exactly $w$ bits are $1$ in each codeword), such that each pair of codewords has Hamming distance at least $d$ (differ in at least $d$ bits).  Used as a block code, it can detect up to $d-1$ errors, and correct up to $\frac{d}{2}-1$ errors with nearest-neighbor decoding.  If the code has $M$ codewords, the code rate is $log_2(M)/n$ bits per transmitted bit.  

$A(n,d,w)$ is the maximum size of a constant weight binary code of length $n$, weight $w$, and distance $d$.  There has been substantial effort on establishing upper and lower bounds on $A(n,d,w)$ for small values of $n$.  Brouwer et al. collected known bounds and established new ones in 1990 \cite{bsss90} and maintain a website tracking subsequent improvements \cite{brouwercodewebsite}.  It has been noted that, at this point, the codes on this website ``are not particularly easy to improve'' \cite{smithpermutation}.  

We focus here on lower bounds, which are established by constructing the largest possible code for each $n,d,w$.  A wide variety of methods have been effective in the past.  These include permutation groups \cite{nurmela,smithpermutation}, constant-weight subsets of linear codes \cite{chee,bsss90}, variations on lexicographic codes \cite{conwaysloane,newtablegreaterthan28}, and randomized search heuristics \cite{MontemanniSmith}.  To help automate the discovery of new constructions, we use CPro1 \cite{cpro1}, a protocol that searches for heuristic strategies for combinatorial constructions.  CPro1 uses Large Language Models (LLMs) to propose diverse strategies and generate programs that implement them for testing.  Here, we use CPro1 to sample and test 4000 methods for constructing constant weight binary codes, finding 2 of these 4000 that succeed in establishing new lower bounds.  With further refinement, these two methods establish 24 new lower bounds for $6 \leq d \leq 18$ and $18 \leq n \leq 35$.  The first of these two methods is a tabu search that operates at the level of bit swaps rather than entire codewords.  The second is a novel greedy heuristic that repeatedly chooses the candidate that maximizes a randomly-scored histogram of distances to previously-added codewords.  We present the new codes and the methods used to obtain them.

\begin{table}[th!]
    \centering
    \begin{tabular}{crrrrl}
       $n$  & $d$ & $w$ & Prior & \textbf{New} & Method \\
         \hline
       18 & 6 & 6 & 132 & \textbf{133} & Bit-Swap Tabu \\[8pt]
       22 & 8 &  9 & 280 & \textbf{292} & RSDH \\
       23 & 8 &  9 & 400 & \textbf{412} & RSDH \\
       23 & 8  & 10  & 616 & \textbf{648} & RSDH \\
       24 & 8 & 9  & 640 & \textbf{656} & RSDH \\
       24 & 8 & 11  & 1288 & \textbf{1378} & S-RSDH $b=1$ \\
       25 & 8  & 7 & 254 & \textbf{255} & S-RSDH $b=1$ \\
       25 & 8  & 9 & 829 & \textbf{837} & RSDH \\
       25 & 8 & 11 & 1662 & \textbf{1702} & S-RSDH $b=1$ \\
       25 & 8 & 12 & 2576 & \textbf{2610} & S-RSDH $b=3$ \\
       26  & 8 & 7 & 257  & \textbf{259} & S-RSDH $b=1$ \\
       26  & 8 & 8 & 760  & \textbf{763} & MS-RSDH $b=2$ \\
       26 & 8 & 11 & 1988 & \textbf{2030} & MS-RSDH $b=2$ \\
       27 & 8 & 10 & 1600 & \textbf{1704} & S-RSDH $b=2$ \\
       28 & 8 & 6 & 130 & \textbf{131} & MS-RSDH $b=2$ \\
       28 & 8 & 10 & 1867 & \textbf{2028} & MS-RSDH $b=3$ \\[8pt] 
       22 & 10 & 8 & 24 & \textbf{25} & Bit-Swap Tabu \\
       25 & 10 & 8 & 48 & \textbf{50} & S-RSDH $b=1$ \\
       28 & 10 & 7 & 37 & \textbf{38} & S-RSDH $b=1$ \\[8pt]
       30 & 12 & 9 & 42 & \textbf{43} & RSDH \\[8pt]
       31 & 16 & 13 & 16 & \textbf{*17}& Bit-Swap Tabu\\
       31 & 16 & 14 & 21 & \textbf{*24} & Bit-Swap Tabu\\
       32 & 16 & 13 & 24 & \textbf{25} & Bit-Swap Tabu\\[8pt]
       35 & 18 & 16 & 21 & \textbf{22}& Bit-Swap Tabu \\
    \end{tabular}
    \caption{New lower bounds on $A(n,d,w)$.  Prior lower bounds are from \cite{brouwercodewebsite}. \textbf{*}~indicates matching upper bound; $A(n,d,w)$ is now exactly determined in these cases.}
    \label{tab:codes}
\end{table}

\section{Main Results}

The refined bit-swap tabu search heuristic is described in Fig.~\ref{fig:bitwisetabupseudocode}.  Methods based on randomly-scored distance histograms are developed further in Section~\ref{sec:rsdh} and described in Figs.~\ref{fig:rsdhpseudocode}-\ref{fig:msrsdhpseudocode}.  These methods were run on each of the open instances (where lower bound is less than upper bound) in the main tables in \cite{brouwercodewebsite} for codes of size less than 3000.  Each method was run on each open instance, on all cores of a 16-core server (AMD 7950X3D), for approximately 2-48 hours.  Larger instances, and instances where early results showed replication of the existing lower bound, were given more time.  It remains possible that even longer runs could have yielded better results with these same methods; this is particularly likely for the largest codes with over 2000 codewords where improvements were still being found late in our runs.

The resulting 24 improved lower bounds are in Table~\ref{tab:codes}, with the method that achieved each result shown.  The full codes, and the programs that generated them, are available on github.\footnote{https://github.com/Constructive-Codes/CWBC}

\section{Prior Work}

LLM-based methods like FunSearch\cite{funsearch}, AlphaEvolve\cite{alphaevolve}, and OpenEvolve\cite{openevolve} have been developed to iteratively evolve programs that implement heuristics or optimizers, and each of these have been applied successfully to generating novel mathematical constructions.  In contrast to this iterative evolution of a set of strategies, the protocol CPro1\cite{cpro1} focuses its computation on sampling large numbers of diverse strategies independently using LLMs, and has been used successfully to solve open instances of combinatorial design problems\cite{jan2025preprint}.  CPro1 has previously generated state-of-the-art binary deletion codes\cite{may2025preprint}, improving on earlier results generated using FunSearch\cite{funsearchdeletioncodes}.  CPro1 has recently aided a successful effort to improve a long-standing upper bound on the number of hyperplanes required to slice all edges of the $n$-dimensional hypercube\cite{edgeslicing}.

Given a textual description of a problem, CPro1\cite{cpro1} uses an LLM to generate diverse candidate strategies, implement them as fast C programs, automatically tune any exposed hyperparameters, and test the candidates on ``Dev'' instances of the problem with known solutions.  Candidate strategies which perform well in solving the Dev instances proceed to testing on Open instances where no solution is yet known.  

\section{Method Development}
\label{sec:methods}

\subsection{Automated Generation of Strategies with CPro1}

\begin{figure}[th!]
\fbox{\parbox{0.97\linewidth}{
\textit{A Constant Weight Binary Code CWBC(n,w,d,s) is a set of at least s vectors of length n over the symbols \{0,1\}, such that (a) each vector has exactly w occurrences of the symbol 1, and (b) each pair of distinct vectors differ in at least d positions.  Given (n,w,d,s), we want to construct a CWBC(n,w,d,s).  For our purposes, n\textless 64, 3\textless w\textless n, d is even and 4\textless=d\textless=20, and s\textless1000.  The code should be output as at least s lines with one vector per line, each vector a space-separated list of elements from \{0,1\}, with the list vectors satisfying constraints (a) and (b).}}}
\caption{The problem definition used with CPro1.}
\label{fig:problemdef}
\end{figure}

\begin{figure}[th!]
\fbox{\parbox{0.97\linewidth}{
\textit{Please suggest 20 different approaches we could implement in C.  \textbf{This is a well-studied problem, and our goal is to go beyond known results, so each of your approaches needs to include a novel element.}  For now, just describe the approaches.  Then I will pick one of the approaches, and you will write the C code to test it.}}}
\caption{CPro1's prompt for listing 20 candidate strategies.  This is appended to the problem definition in Fig.~\ref{fig:problemdef}.  Half the runs here included the \textbf{bold} sentence of the prompt requesting strategies with a novel element; the other half use the default configuration of CPro1 which omits this extra sentence.}
\label{fig:strats}
\end{figure}

\begin{table}[th!]
    \centering
    \setlength{\tabcolsep}{5.4pt} 
    \begin{tabular}{rrrr|rrrr|rrrr|rrrr}
        n & d & w & s & n & d & w & s & n & d & w & s & n & d & w & s \\
        \hline
        10 & 4  & 4  & 30  & 13 & 6  & 5  & 18  & 14 & 6  & 6  & 42  & 15 & 6  & 7  & 69  \\
        18 & 8  & 6  & 21  & 20 & 8  & 7  & 80  & 17 & 6  & 7  & 149 & 17 & 6  & 8  & 166 \\
        18 & 6  & 6  & 119 & 19 & 6  & 6  & 155 & 20 & 6  & 8  & 528 & 21 & 6  & 8  & 698 \\
        27 & 6  & 5  & 234 & 22 & 8  & 9  & 252 & 23 & 8  & 6  & 69  & 24 & 8  & 6  & 70  \\
        17 & 6  & 7  & 166 & 17 & 6  & 8  & 184 & 18 & 6  & 6  & 132 & 19 & 6  & 6  & 172 \\
        20 & 6  & 8  & 588 & 21 & 6  & 8  & 775 & 27 & 6  & 5  & 260 & 22 & 8  & 9  & 280 \\
        23 & 8  & 6  & 77  & 24 & 8  & 6  & 78  & 28 & 8  & 6  & 130 & 21 & 10 & 9  & 27  \\
        21 & 10 & 10 & 38  & 22 & 10 & 8  & 24  & 22 & 10 & 9  & 35  & 24 & 10 & 8  & 38  \\
    \end{tabular}
    \caption{The 32 Dev instances used with CPro1.  For each of these, a code with $n,d,w$ having $s$ codewords is known to exist.  Candidate CPro1 programs are scored based on their ability to produce a code with these parameters.}
    \label{tab:devinstances}
\end{table}

\begin{table}[th!]
    \centering
    \setlength{\tabcolsep}{5.4pt} 
    \begin{tabular}{rrrr|rrrr|rrrr|rrrr}
        n & d & w & s & n & d & w & s & n & d & w & s & n & d & w & s \\
        \hline
        17 & 6  & 7  & 167 & 17 & 6  & 8  & 185 & 18 & 6  & 6  & 133 & 19 & 6  & 6  & 173 \\
        20 & 6  & 8  & 589 & 21 & 6  & 8  & 776 & 27 & 6  & 5  & 261 & 22 & 8  & 9  & 281 \\
        23 & 8  & 6  & 78  & 24 & 8  & 6  & 79  & 28 & 8  & 6  & 131 & 21 & 10 & 9  & 28  \\
        21 & 10 & 10 & 39  & 22 & 10 & 8  & 25  & 22 & 10 & 9  & 36  & 24 & 10 & 8  & 39  \\
    \end{tabular}
    \caption{The 16 Open instances used with CPro1.  The programs scoring best on the Dev instances are then tested on these Open instances.}
    \label{tab:openinstances}
\end{table}

For our problem, the textual definition used with CPro1 is shown in Fig.~\ref{fig:problemdef}, the Dev instances are listed in Table~\ref{tab:devinstances} and the Open instances used in the CPro1 runs are listed in Table~\ref{tab:openinstances}. In these experiments, CPro1 uses OpenAI o4-mini \cite{o4mini} as the LLM.  

CPro1 initially prompts the LLM to generate a list of 20 strategies, and then implements and tests each one.  Here, a single run consists of 10 batches of 20 for a total of 200 strategies per run.  We performed 20 such runs, so 4000 strategies were tested. 

CPro1's initial strategy-listing LLM prompt is given in~\ref{fig:strats}.  CPro1's original strategy-listing prompt\cite{cpro1} omits the boldface sentence requesting that each strategy include a novel element.  Half our runs use the original form, and the other half test the new prompt by including the sentence requesting a novel element.

Among the 2000 strategies generated without the novelty prompt, 33.5\% solved at least one Dev instance.  This fell to 21.5\% when using the novelty prompt; it is apparently more risky to request novel strategies.  However the best two strategies with the novelty prompt solved 18 and 17 of the 32 Dev instances, better than the 16 Dev instances solved without the novelty prompt.  The best two strategies generated with the novelty prompt each solved one Open instance; no Open instances were solved without the novelty prompt.

\begin{figure}[tb]
\centering
\begin{minipage}{1\linewidth}
\begin{tcolorbox}[pseudo/ruled]
\begin{pseudo*}
Conflict-Pairs($Code$): \textbf{return} $\{$distinct $(x,y)\in Code$ with \textbf{HD}$(x,y)$$<$$d\}$\\
\\
Penalty($Code$): \textbf{return} $\sum_{(x,y)\in\textrm{Conflict-Pairs}(Code)} d-$\textbf{HD}$(x,y)$ \\-
\\
Bit-Swap-Tabu($n,w,d,s$):\\+
  $Code = \{s $ distinct random $n$-bit words of weight $w \}$ // Initialize \\
  $minP$ = Penalty($Code$) \\
  \textbf{while} Penalty$(Code)>0$: \\+
    $(x,y)$ = random pair from Conflict-Pairs$(Code)$\\
    \textbf{for} $(c,o) \in \{(x,y),(y,x)\}$: \\+
      $p_{b} = \{$bit positions where both $c$ and $o$ are $b\}$, for $b \in \{0,1\}$ \\ 
      \textbf{for} $p\in p_{1}, q\in p_{0}$: \\+
        $P_{\textrm{move}} =$ Penalty from clearing bit $p$ and setting bit $q$ in $c$  \\
        \textbf{if} (move is tabu) \& ($P_{\textrm{move}} \geq minP)$: \textbf{skip} // tabu, not aspiration \\
        \textrm{if} $P_{\textrm{move}}<best$: $best = P_{\textrm{move}}$; record $c,p,q$ that obtain $best$ \\--
    \textbf{Update} $Code$ with recorded best move $c,p,q$ \\
    \textbf{Set} tabu: forbid setting $p$\&clearing $q$ in $c$ for \textbf{rand\_int}($t_{min},t_{max}$) steps\\
    \textbf{if} $best<minP$: $minP=best$ // track global best \\
    \textbf{if} no valid move \textbf{or} no improvement for max\_no\_improve\_restart steps: \\+
      $Code = \{s $ distinct random $n$-bit words of weight $w \}$ // re-initialize \\
      $minP$ = Penalty($Code$) \\--
  \textbf{return} $Code$ \\
\end{pseudo*}
\end{tcolorbox}
\end{minipage}
\caption{Bit-Swap Tabu Search.  In the efficient implementation, Conflict-Pairs and Penalty are maintained incrementally.  Note \textbf{HD}($x,y$) is the Hamming Distance between binary words $x$ and $y$.  We set $t_{min}=5$, $t_{max}=15$, max\_no\_improve\_restart=1M.   }
\label{fig:bitwisetabupseudocode}
\end{figure}

The first successful strategy is a tabu search heuristic, that operates at the level of bit-swapping rather than adding and removing entire codewords.  During the CPro1 run, for $n=22$ $d=10$ $w=8$ it found 25 codewords, improving upon the previous lower bound of 24.  After simplifying the implementation by hand and running it further, it generated 6 of the new lower bounds in Table~\ref{tab:codes}.  This includes two lower bounds that exactly match the upper bound, exactly determining $A(31,16,13)=17$ and $A(31,16,14)=24$.  The algorithm is shown in Fig.~\ref{fig:bitwisetabupseudocode}.

\begin{figure}[t]
\centering
\begin{minipage}{1\linewidth}
\begin{tcolorbox}[pseudo/ruled]
\begin{pseudo*}
Rand-Vec($d$,$w$): // Returns $SV$, a random Scoring Vector \\+ 
  \textbf{For} $x \in \{d,d+2,\ldots,2w\}$: $SV[x]$ = \textbf{rand\_int}($-2^{31},2^{31}$) \\
  \textbf{while} $SV[d] < \max_{x} SV[x]$: \\+
    $SV[d]$ = \textbf{rand\_int}($0,2^{31}$) // distance $d$ gets highest score \\-
  \textbf{return} $SV$ \\
\\-
Greedy-Build($d$,$w$,$Code$,$Candidates$,$Score$): // populates $Code$ \\+
  $SV$ = Rand-vec($d$,$w$) \\
  \textbf{while} $Candidates$ \textbf{is not empty}: \\+
    $M = \argmax_{C \in Candidates} Score[C]$ // \textbf{\textit{break ties randomly}} \\
    \textbf{Append} $M$ \textbf{to} $Code$ \\
    \textbf{For each} $C \in Candidates$: \\+
      \textbf{if} \textbf{HD}$(C,M)<d$: \textbf{Delete} $C$ \textbf{from} $Candidates$ \\
      \textbf{else}: $Score[C]$ += $SV$[\textbf{HD}($C,M$)] \\-
\\--
RSDH($n,w,d,s$):\\+
  \textbf{repeat until} $|Code|\geq s$: \\+
    $Code = \{\}$ \\
    $Candidates = \{n$-bit words of weight $w\}$ \\
    \textbf{For each} $C \in Candidates: Score[C] = 0$ \\
    Greedy-Build($d,w,Code,Candidates,Score$) \\-
  \textbf{return} $Code$ \\ 
\end{pseudo*}
\end{tcolorbox}
\end{minipage}
\caption{Random-Score Distance Histograms (RSDH).  $n,w,d$ are the parameters of the code, and $s$ is the target size (number of codewords).  Note \textbf{HD}($x,y$) is the Hamming Distance between binary words $x$ and $y$.}
\label{fig:rsdhpseudocode}
\end{figure}

\begin{figure}
\centering
\begin{minipage}{1\linewidth}
\begin{tcolorbox}[pseudo/ruled]
\begin{pseudo*}
S-RSDH($n,w,d,s$,$b$,$t$):\\+
  \textbf{repeat until} $|Code|\geq s$: \\+
    $Code1 = \{\}$ \\
    $r$ = random $b$-bit integer \\
    $Candidates = \{n$-bit words of weight $w$ such that last $b$ bits are $r\}$ \\
    \textbf{For each} $C \in Candidates: Score[C] = 0$ \\
    Greedy-Build($d,w,Code1,Candidates,Score$) // first slice \\
    \\
    \textbf{repeat} $t$ \textbf{times}: // take best of $t$ completions \\+
      $Code2$ = $Code1$ \\
      $SV$ = Rand-Vec($d$,$w$) \\
      $Candidates = \{n$-bit words of weight $w\}$ \\
      \textbf{For each} $C \in Candidates$: \\+
        \textbf{if} $\exists x \in Code2$ with \textbf{HD}$(C,x)$$<$$d$: \textbf{Delete} $C$ \textbf{from} $Candidates$\\
        \textbf{else} $Score[C] = \sum_{x \in Code2} SV[$\textbf{HD}$(C,x)]$ \\-
      Greedy-Build($d,w,Code2,Candidates,Score$) // remainder of code \\
      \textbf{if} $|Code2|>|Code|$: $Code = Code2$ // $Code$ is best of $t$ \\--
  \textbf{return} $Code$ \\ 
\end{pseudo*}
\end{tcolorbox}
\end{minipage}
\caption{Sliced Random-Score Distance Histograms (S-RSDH).  $b$ is the number of bits that define the slice; we use $b \in \{1,2,3\}$. $t$ is the number of trials for the remainder of the code; we fix $t=1000$.  Greedy-Build, Rand-Vec, and \textbf{HD} are defined in Fig.~\ref{fig:rsdhpseudocode}.}
\label{fig:srsdhpseudocode}
\end{figure}

\begin{figure}
\centering
\begin{minipage}{1\linewidth}
\begin{tcolorbox}[pseudo/ruled]
\begin{pseudo*}
MS-RSDH($n,w,d,s$,$b$):\\+
  \textbf{repeat until} $|Code|\geq s$: \\+
    $Code = \{\}$ \\
    $R$ = random shuffle of $\{0,1,...,2^{b}-1\}$ \\
    $trials = 1$ // first slice gets 1 trial, remainder get $t$ trials \\
    \textbf{for} {$r \in R$}: \\+
      $Code1 = Code$ \\
      \textbf{repeat} $trials$ \textbf{times}: // take best result of $trials$ for this slice \\+
        $Code2$ = $Code1$ \\
        $Candidates = \{n$-bit words of weight $w$ such that last $b$ bits are $r\}$ \\
        $SV$ = Rand-Vec($d$,$w$) \\
        \textbf{For each} $C \in Candidates$: \\+
          \textbf{if} $\exists x \in Code2$ with \textbf{HD}$(C,x)$$<$$d$: \textbf{Delete} $C$ \textbf{from} $Candidates$\\
          \textbf{else} $Score[C] = \sum_{x \in Code2} SV[$\textbf{HD}$(C,x)]$ \\-
        Greedy-Build($d,w,Code2,Candidates,Score$) \\
        \textbf{if} $|Code2|>|Code|$: $Code = Code2$ // $Code$ is best of $t$ \\-
      $trials = t$ // $t$ trials after first slice \\--
  \textbf{return} $Code$ \\
\end{pseudo*}
\end{tcolorbox}
\end{minipage}
\caption{Multi-Sliced Random-Score Distance Histograms (MS-RSDH).}
\label{fig:msrsdhpseudocode}
\end{figure}

The second successful strategy was proposed by CPro1 as a reinforcement learning method, using a distance histogram representation for learned policy.  During the CPro1 run, for $n=22$ $d=8$ $w=9$ it found 281 codewords, improving upon the previous lower bound of 280.  This strategy was also distinguished in that it was the only method which solved Dev instances with over 200 codewords; other proposed methods were only effective for smaller sets of codewords.  The proposed ``reinforcement learning'' method adds one codeword at a time, choosing the candidate that maximizes the dot product of a scoring vector (the policy) with the histogram of distances between the candidate and the previously-added codewords, breaking ties randomly.  We find that the reinforcement learning component turns out to be unnecessary.  There is no positive reinforcement, since the algorithm terminates when reaching the target $s$ and only applies negative reinforcement until then.  The negative reinforcement ``learning'' simply has the scoring vector take a random walk.  Further experimentation shows the random walk could be replaced by randomly reinitializing the scoring vector before each greedy construction.  The resulting method, \textit{Random-Score Distance Histograms (RSDH)}, is shown in Fig.~\ref{fig:rsdhpseudocode}.  The method is analyzed and developed further in the next section. 

\subsection{Random-Score Distance Histograms (RSDH)}
\label{sec:rsdh}

Running RSDH on additional open instances (beyond those listed in Table~\ref{tab:openinstances}) identifies several codes that improve over existing lower bounds.  We find different scoring vectors are needed for different parameters ($n$,$w$,$d$).  However, successful scoring vectors consistently place the highest value on minimum distance $d$, which makes sense as this prefers codewords which are packed tightly together.  Therefore, in Fig.~\ref{fig:rsdhpseudocode} we restrict to scoring vectors that score $d$ highest.

\begin{figure}
    \centering
    \includegraphics[width=1\linewidth]{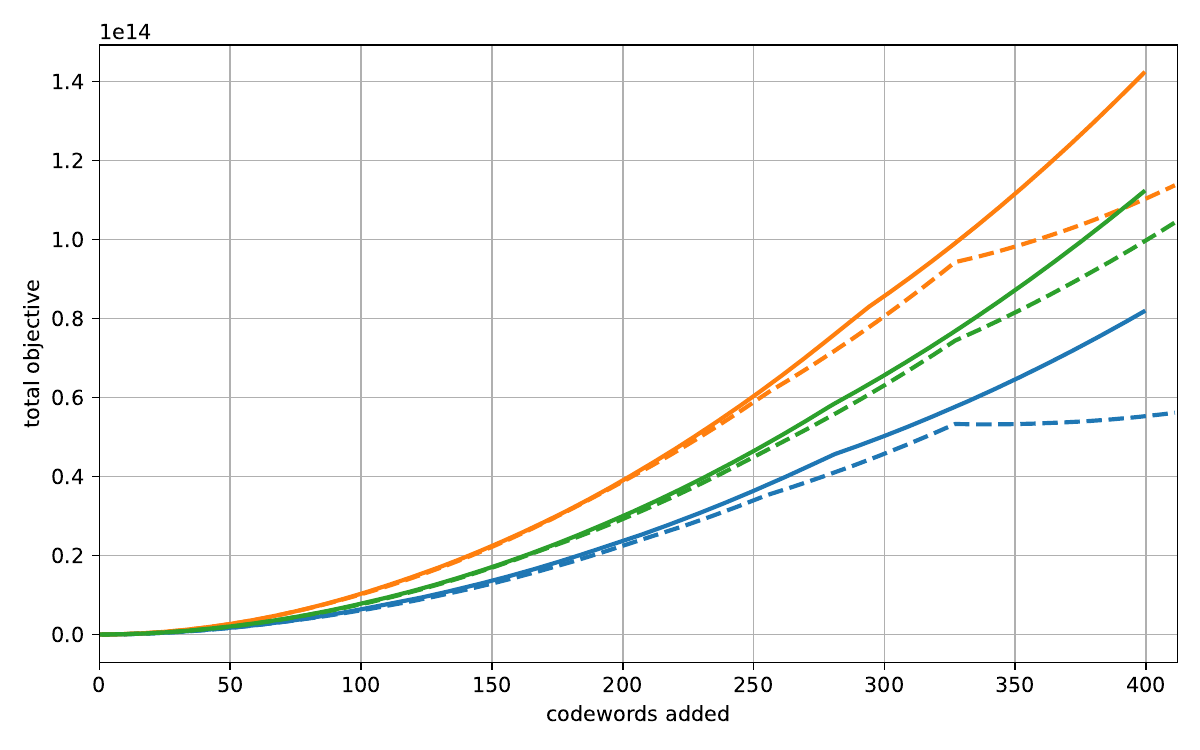}
    \caption{Example trajectories of total objective function for $n=23$ $d=8$ $w=9$.  Each color uses a different scoring vector.  The solid trajectory reaches 400, and the dashed trajectory reaches 412 with the same scoring vector.}
    \label{fig:traj}
\end{figure}

The random tie-breaking in Fig.~\ref{fig:rsdhpseudocode} is essential.  As an example, the previous lower bound for $n=23$ $d=8$ $w=9$ was 400 codewords; here, we find 412.  However, across several scoring vectors that can produce 412, rerunning with the same scoring vector yields 412 only about 0.5\% of the time.  More common is 400, which is produced about 14\% of the time with the same scoring vectors.  Example trajectories are shown in Figure~\ref{fig:traj}.  We define the total objective function as the dot product of the scoring vector with the histogram of pairwise distances; the trajectory shows how this objective function changes as codewords are added.

\begin{figure}
    \centering
    \includegraphics[width=1\linewidth]{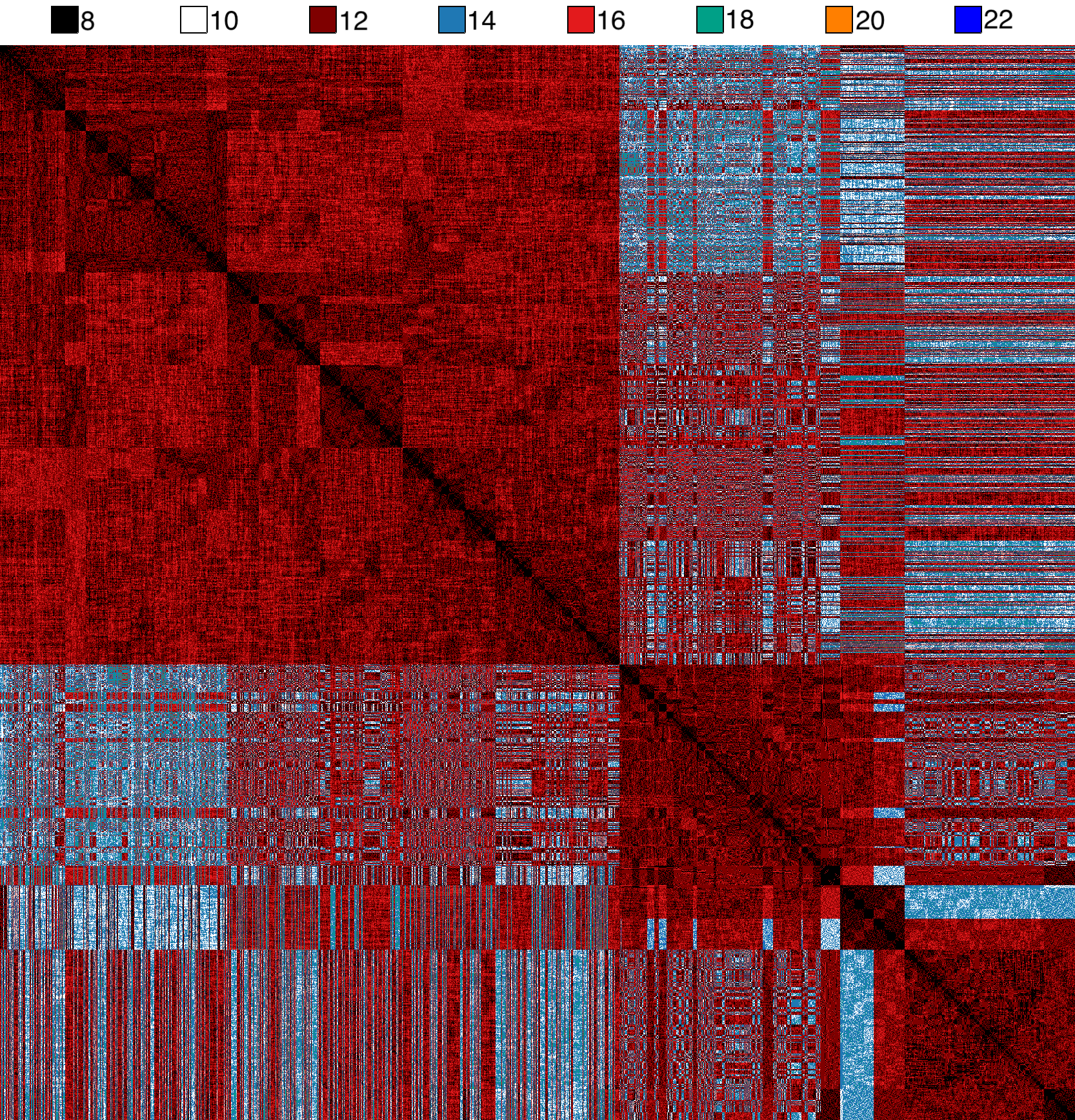}
    \caption{Distance map for a $n=25$ $d=8$ $w=11$ code with 1666 codewords generated by RSDH.  The 1666 codewords are ordered in the sequence chosen by the greedy algorithm, from left to right and top to bottom.  Row $i$ column $j$ is colored according to the distance between codeword $i$ and codeword $j$, as shown in the color key at the top of the figure.  The red/maroon/black upper-left block is size 960x960.}
    \label{fig:distmap}
\end{figure}

The method in Fig.~\ref{fig:rsdhpseudocode} could be considered to be an approximate greedy optimization of the final total objective function. This raises the question of whether better optimization (rather than a single greedy construction) could yield better codes.  However, Figure~\ref{fig:traj} shows that for various scoring vectors that produce 412 codewords for $n=23$ $d=8$ $w=9$, trajectories that produce 400 actually achieve a higher objective than trajectories that produce 412.  So, better optimization here would select for trajectories that produce a worse result; it is important that we stay with the greedy optimizer with random tie-breaking.  Note that the dropoff in slope towards the end of the 412 trajectories is due to late introduction of low-scoring large distances in these trajectories; these are distances that do not occur in the 400 trajectories.

RSDH obtains 1666 codewords for $n=25$ $d=8$ $w=11$, improving on the previous best of 1662 codewords. The construction of this previous best, like several others with $d=8$, was based on constant-weight subsets of the Golay code \cite{bsss90}.  So, what structure can we find in our 1666?  Figure~\ref{fig:distmap} shows the distance between each pair of codewords in the 1666.  This shows that the first 960 codewords are all distance 8, 12, or 16 from one another, with other distances only appearing after that.  Looking at the codewords, we see the first 960 codewords have a constant column that is always 1.  Removing this column, these 960 codewords turn out to be isomorphic to the best-known $n=24$ $d=8$ $w=10$, which is derived from the 23-bit Golay code by combining a 560-codeword subset of weight 10, with a 400-codeword subset of weight 9 with a 1 appended, yielding 960 codewords.  So, Fig.~\ref{fig:rsdhpseudocode} is able to construct new codes that are based on linear codes like the Golay code.

\subsubsection{Sliced Random-Score Distance Histograms}

A single scoring vector is doing a lot of work in Fig.~\ref{fig:distmap}, yielding both the initial 960 codewords with a constant column, and then the extension which introduces new distances.  It raises the question whether using different scoring vectors for different slices of the code may be productive.  

A potential alternative approach to $n=25$ $d=8$ $w=11$ would be: constrain one bit to a constant, and use a random scoring vector to produce a maximal set of codewords on this ``slice'' of the code.  Then remove the constraint and complete the code: initialize the score of the remaining allowed candidates with a new second random scoring vector, and then choose a final third random scoring vector that is applied to the distances among the new unconstrained codewords.  Choosing the second and third scoring vectors independently appears to work better than making them equal.

Figure~\ref{fig:srsdhpseudocode} describes this ``sliced'' random-score distance histograms (S-RSDH) approach, generalized from constraining one bit to constraining $b$ bits; we test $b \in \{1,2,3\}$.  For each result on the first slice, we sample 1000 completions of the code; these completions are fast since the number of remaining candidates is relatively small after the first slice is completed.  

S-RSDH yields improved results for some parameters, including $n=25$ $d=8$ $w=11$ with $b=1$ where we obtain 1702 codewords.  The 1702's initial slice has the last bit constrained to 0, and obtains 1218 codewords for this slice, and then adds 484 codewords with the last bit 1 to complete the code.  Note that, while the initial 1218 is a maximal code (no codeword can be added within the slice), is not optimal for $n=24$ $d=8$ $w=11$ (1288 was already known, and results in this paper show 1378 is possible).  

For $b>1$, we can also divide into $2^b$ slices, one for each setting to the $b$ bits.  Figure~\ref{fig:msrsdhpseudocode} randomly orders the $2^b$ slices, testing 1000 candidates in each and greedily choosing the biggest result before continuing to the next slice.  This ``multi-sliced'' random-score distance histograms (MS-RSDH) performs well for some parameters.

Table~\ref{tab:codes} has the improved lower bounds from these methods.   

\section{Conclusion}

We have established 24 new lower bounds on constant weight binary codes by constructing improved larger codes, for distances $6 \leq d \leq 18$ and bit sizes $18 \leq n \leq 35$.  The novel Random-Score Distance Histograms approach, and variants of it, established the majority of the new lower bounds.  The heuristics we used were based on a sample of 4000 strategies using CPro1.  There is a random element to this automated discovery: it seems likely that different samples from CPro1 would find different heuristics (which could be more or less successful), and significantly scaling up the sample size might find even more successful heuristics.
\bibliographystyle{abbrvnat}
\bibliography{cwbc}

@online{o4mini,
      title={Open{A}{I} o3 and o4-mini System Card}, 
      author={Open{A}{I}},
      year={2025},
      note={https://cdn.openai.com/pdf/2221c875-02dc-4789-800b-e7758f3722c1/o3-and-o4-mini-system-card.pdf} 
}

@article{conwaysloane,
  title={Lexicographic codes: error-correcting codes from game theory},
  author={Conway, John and Sloane, N},
  journal={IEEE Transactions on Information Theory},
  volume={32},
  number={3},
  pages={337--348},
  year={2003}
}

@article{bsss90,
  title={A new table of constant weight codes},
  author={Brouwer, Andries E and Shearer, James B and Sloane, Neil JA and Smith, Warren D},
  journal={IEEE Transactions on Information Theory},
  volume={36},
  number={6},
  pages={1334--1380},
  year={1990}
}

@article{newtablegreaterthan28,
  title={A new table of constant weight codes of length greater than 28},
  author={Smith, Derek H and Hughes, Lesley A and Perkins, Stephanie},
  journal={The electronic journal of combinatorics},
  pages={A2},
  year={2006}
}

@misc{brouwercodewebsite,
  author = "Andries E. Brouwer",
  title = "Bounds for binary constant weight codes",
  note="https://aeb.win.tue.nl/codes/Andw.html"
}

@inproceedings{cpro1,
  title={L{L}{M}-Generated Search Heuristics Can Solve Open Instances of Combinatorial Design Problems},
  author={Rosin, Christopher D},
  booktitle={The 5th Workshop on Mathematical Reasoning and AI at NeurIPS},
  year={2025}
}

@article{montemannismith,
  title={Heuristic algorithms for constructing binary constant weight codes},
  author={Montemanni, Roberto and Smith, Derek H},
  journal={IEEE Transactions on Information Theory},
  volume={55},
  number={10},
  pages={4651--4656},
  year={2009}
}

@article{smithpermutation,
  title={Some constant weight codes from primitive permutation groups},
  author={Smith, Derek H and Montemanni, Roberto},
  journal={The electronic journal of combinatorics},
  volume={19},
  number={4},
  year={2012}
}

@article{nurmela,
  title={New constant weight codes from linear permutation groups},
  author={Nurmela, Kari J and Kaikkonen, Markku K and Ostergard, PRJ},
  journal={IEEE Transactions on Information Theory},
  volume={43},
  number={5},
  pages={1623--1630},
  year={1997}
}

@article{chee,
  title={New constant-weight codes from propagation rules},
  author={Chee, Yeow Meng and Xing, Chaoping and Yeo, Sze Ling},
  journal={IEEE Transactions on Information Theory},
  volume={56},
  pages={1596--1599},
  year={2010}
}

@article{funsearch,
  title={Mathematical discoveries from program search with large language models},
  author={Romera-Paredes, Bernardino and Barekatain, Mohammadamin and Novikov, Alexander and Balog, Matej and Kumar, M Pawan and Dupont, Emilien and Ruiz, Francisco JR and Ellenberg, Jordan S and Wang, Pengming and Fawzi, Omar and others},
  journal={Nature},
  volume={625},
  number={7995},
  pages={468--475},
  year={2024}
}

@misc{alphaevolve,
  title={Alpha{E}volve: A coding agent for scientific and algorithmic discovery},
  author={Novikov, Alexander and Vu, Ngan and Eisenberger, Marvin and Dupont, Emilien and others},
  note={arXiv:2506.13131},
  year={2025}
}

@misc{openevolve,
  author = "Asankhaya Sharma",
  title = "Open{E}volve",
  note="https://github.com/algorithmicsuperintelligence/openevolve"
}

@misc{jan2025preprint,
  title={Using Code Generation to Solve Open Instances of Combinatorial Design Problems},
  author={Rosin, Christopher D.},
  note={arXiv:2501.17725},
  month={January},
  year={2025}
}

@misc{may2025preprint,
  title={Using Reasoning Models to Generate Search Heuristics that Solve Open Instances of Combinatorial Design Problems},
  author={Rosin, Christopher D.},
  note={arXiv:2505.23881},
  month={May},
  year={2025}
}

@misc{funsearchdeletioncodes,
  title={L{L}{M}-Guided Search for Deletion-Correcting Codes},
  author={Weindel, Franziska and Heckel, Reinhard},
  note={arXiv:2504.00613},
  year={2025}
}

@article{edgeslicing,
  title={Improved Upper Bounds for Slicing the Hypercube},
  author={Soiffer, Duncan and Itty, Nathaniel and Rosin, Christopher D and Bruell, Blake and DiCicco, Mason and S{\'a}rk{\"o}zy, G{\'a}bor N and Offstein, Ryan and Reichman, Daniel},
  journal={arXiv preprint arXiv:2602.16807},
  year={2026}
}

\end{document}